# Phase Noise Tolerance for Low-Pilot-Overhead OFDM Terahertz Links Beyond 64-QAM

Bowen Liu*
*Department of Electronics and Electrical Engineering*
*Keio University*
Yokohama, Japan
b.liu@phot.elec.keio.ac.jp

Takasumi Tanabe
*Department of Electronics and Electrical Engineering*
*Keio University*
Yokohama, Japan
takasumi@elec.keio.ac.jp

***Abstract*—THz wireless communications have garnered significant attention due to their unprecedented data rates enabled by the abundant untapped spectrum. However, advanced modulation formats beyond 64-QAM remain largely unexplored, as phase errors introduced during up/down-conversion severely limit system performance. Particularly, OFDM transmission is highly susceptible to aggravated ICI induced by phase noise, undermining the orthogonality of subcarriers. While PLLs and pilot-assisted compensation can mitigate phase errors, excessive pilot overhead compromises spectral efficiency and energy consumption, and white phase noise remains unrecoverable. Therefore, quantifying phase noise tolerance is essential for practical physical layer protocols. Here, we reveal the impact of phase noise in a 64-QAM, 2048-subcarrier OFDM THz transmission system. 3σ-error estimation is proposed to quantify phase noise tolerance, indicating an intuitive EVM threshold of approximately 5%. This threshold further delineates the trade-offs among phase noise levels, SNR requirements, and pilot overhead. Moreover, by benchmarking representative oscillators with distinct phase noise spectra, microring resonators (MRRs) are identified as indispensable enablers for low-pilot-overhead OFDM THz links operating beyond 64-QAM.**

## I. Introduction

The rapid expansion of the digital economy is driving an unprecedented demand for mobile data access, laying the foundation for high-speed wireless links that leverage advanced modulation formats and enhanced channel coding schemes [1, 2]. In 2022, monthly data traffic in the Asia-Pacific region increased by 11 Gb per user, accompanied by the addition of approximately 3.1 billion new mobile subscribers [3]. However, despite this surge, approximately 47% of wireless connections in the region still operate below 100 Mbps peak data rates, with an average access speed of merely 28.5 Mbps [3]. In contrast, Terahertz (THz) wireless communications offer transmission speeds that rival, and in certain scenarios surpass, those of optical fibers, particularly in short-range applications [4, 5]. The 300-GHz transmission window alone provides up to 44 GHz of underutilized bandwidth, which is approximately 110 times wider than the spectrum allocated in 5G NR standards [6]. Significant advancements have been demonstrated in this field over the past decade. In 2011, a 100 Gbit/s 16-QAM wireless link was achieved in the 75–110 GHz band over a 1.2-m air channel [7]. Subsequent works extended data rates to 160 Gbit/s within the 300–500 GHz band by employing eight channels spaced at 25 GHz intervals [8]. Robust THz sources enabled further demonstrations of 32-QAM wireless transmission, including a 100 Gbit/s link reported by L. John *et al*. in 2020 [9], and a 160 Gbit/s link using self-injection-locked microcombs by C. Koos *et al*. in 2023 [10], which was later enhanced to 250 Gbit/s [11]. T. Nagatsuma *et al*. reported a 220 Gbit/s 32-QAM free-space transmission over 214 m [12]. Additionally, quantum-dash lasers with external coherent feedback achieved a remarkable 12 Tbit/s 32-QAM THz link [13]. Moreover, progress towards higher-order modulation such as 64-QAM, has also been demonstrated [14-17]. In 2023, Tetsumoto *et al*. proposed a noise-reduction technique for 64-QAM transmission by simultaneously modulating multiple comb lines to enhance the optical signal-to-noise ratio (OSNR) of THz carriers [14]. A 210 Gbit/s 64-QAM link was experimentally validated [15], while an aggregate throughput approaching 1 Tbit/s was reported in 2024 by utilizing four-channel carrier aggregation combined with advanced multiplexing strategies [16]. Meanwhile, a 64-QAM OFDM transmission achieving approximately 613 Gbit/s has also been demonstrated [17].

However, the error performance of THz transmission remains far from practical application. This is mainly due to the increasing sensitivity of higher-order modulations such as 64-QAM and beyond, to phase noise impairments. As constellation points become denser, the decision threshold tightens, making symbols at the outer edges extremely vulnerable to phase disturbances introduced during up/down-conversion. Phase noise originates from the inherent phase uncertainty of oscillators, manifesting as random jitter in the time domain and resulting in spectral broadening around the carrier frequency in the frequency domain as explained in Fig. 1 [18]. The phase noise level is commonly characterized by measuring the power spectral density (PSD) within a 1-Hz bandwidth at a given frequency offset from the carrier, expressed in dBc/Hz. The overall phase noise spectrum typically exhibits colored noise characteristics, including flicker FM ($1/f^3$), white FM ($1/f^2$), flicker PM ($1/f$), and white PM components, as described by the Leeson model [18, 19]. Key parameters include the noise figure ($F$), thermal noise ($kT$), oscillator power ($P_s$), and the loaded quality factor ($Q$).

$$L(f) = 10log_{10}[\frac{2FkT}{P_s}(1 + (\frac{f_{LO}}{2Qf})^2)(1 + \frac{f_c}{f})] \tag{1}$$

Therefore, high-Q resonators such as microring resonators (MRRs) can effectively suppress phase noise by shifting the corner frequency ($f_{LO}/2Q$) closer to the carrier and narrowing the oscillator linewidth. While low-frequency components can be mitigated through phase-locking loops (PLLs) and pilot-assisted estimation [20], high-frequency components characterized as white phase noise remain irrecoverable. This instantaneous phase noise ($\varphi_{inst}$) has often been overlooked, yet it induces local rotation and diffusion errors on constellation points that cannot be compensated as depicted in Fig. 1 (e). The impact is even more detrimental in OFDM systems, where phase noise destroys the orthogonality of subcarriers, causing energy leakage and inter-carrier interference (ICI). This mechanism converts phase noise into amplitude distortions, thereby exacerbating error vector magnitude (EVM) in conjunction with thermal noise. Unlike electronic oscillators, photonic approaches suppress noise at the optical carrier and benefit from a 20-dB/dec frequency division advantage when down-converted to microwaves. MRRs leverage this frequency division while maintaining an ultra-high-Q factor [21], making them indispensable for achieving low-noise, low-pilot-overhead OFDM THz links beyond 64-QAM.

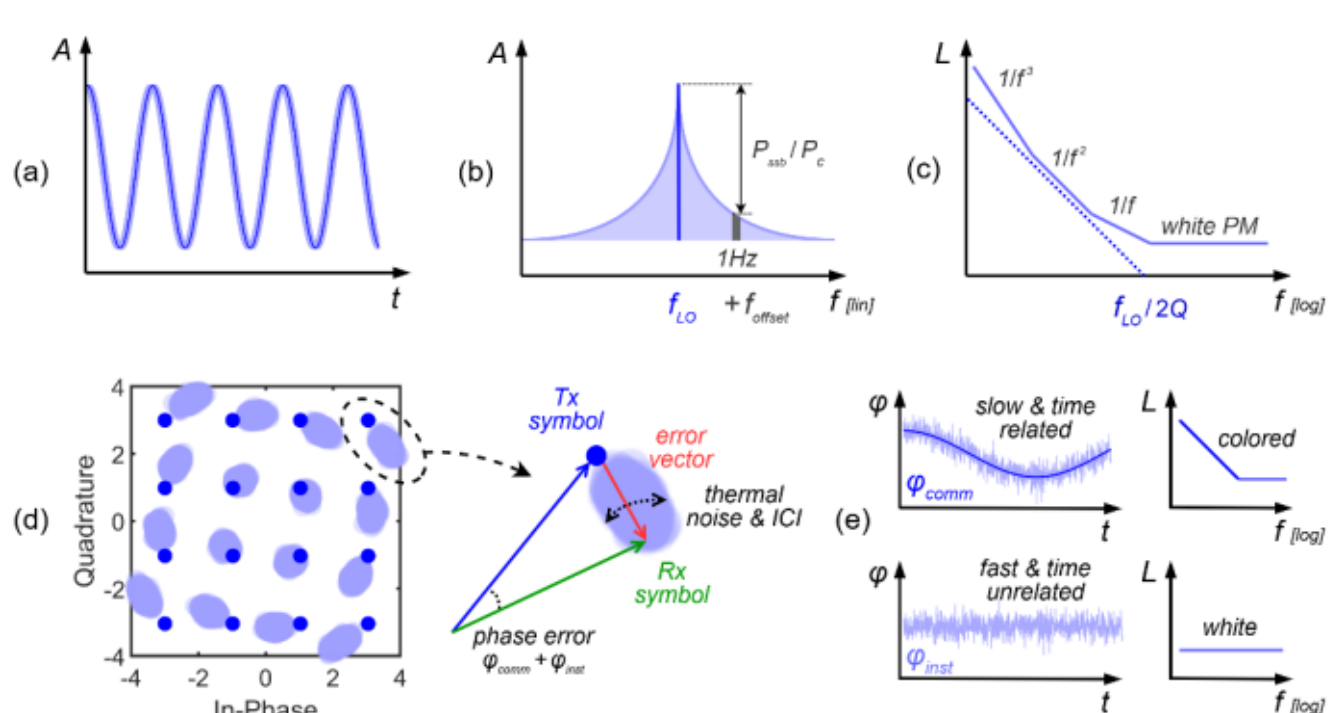

Fig. 1. Phase noise in the (a) time domain and (b) frequency domain; (c) PSD reveals different colored noise contributions; (d) Phase noise rotation error in the constellation diagram can be estimated via EVM; (e) Common and instantaneous components contribute differently to the phase errors.

Yet a quantitative understanding of phase noise, particularly its instantaneous component, on system performance remains unaddressed. In OFDM systems employing modulations beyond 64-QAM, the trade-off between pilot overhead and system performance becomes increasingly critical, as instantaneous phase noise induces severe impairments that standard compensation techniques cannot fully mitigate. To enable low-pilot-overhead OFDM protocols in the 300-GHz band, this work models phase noise based on experimental spectra. We quantify the error-free tolerance of 64-QAM codewords without forward error correction (FEC) by applying 3σ-error estimation, which directly reveals the threshold relationships among phase noise, signal-to-noise ratio (SNR), and pilot overhead. Our analysis highlights that instantaneous phase noise imposes a fundamental limit on performance ceiling, which cannot be recovered through conventional phase tracking or pilot equalization. As a solution, stabilized MRR referenced to ultra-high-Q cavities, emerge as a key enabler for achieving low-noise, low-pilot-overhead OFDM THz links beyond 64-QAM. The findings provide critical insights for the development of physical-layer protocols in future THz wireless communications.

This work was supported by JST, CRONOS, Japan (JPMJCS24N7).

## II. PRINCIPLE

### A. 64-QAM 2048-OFDM Terahertz Links

The schematic diagram of modeled OFDM THz wireless transmission system is illustrated in Fig. 2. A pseudo-random bit stream is first encoded in the baseband processor and mapped onto a 64-QAM constellation. The modulated symbols are then fed into a 2048-point OFDM processor, where serial-to-parallel (S/P) conversion is performed to distribute the symbols across 2048 sub-carriers. Pilot tones are inserted using a comb-type arrangement for channel estimation and phase noise compensation. Following pilot insertion, an inverse fast Fourier transform (IFFT) is applied to generate the time-domain OFDM signal. The sub-carrier spacing (SCS) is set to 30 kHz, resulting in 14 OFDM symbols every time slot with 0.5-ms duration. A cyclic prefix (CP) equivalent to 16 sub-carriers is appended to each OFDM symbol to mitigate inter-symbol interference (ISI) caused by multipath fading. The parallel data is then serialized and passed to a digital-to-analog converter (DAC) for up-conversion to the THz band. Phase noise is introduced during up-conversion. The up-converted signal is transmitted via a lensed horn antenna and propagates through a wireless channel modeled as additive white Gaussian noise (AWGN) combined with a 3-tap finite impulse response (FIR) fading filter. At the receiver, the signal is down-converted to the intermediate frequency, filtered, and digitized by an analog-to-digital converter (ADC). After subsequent digital signal processing (DSP), the system performance is evaluated in EVM, quantifying the overall degradation caused by phase noise.

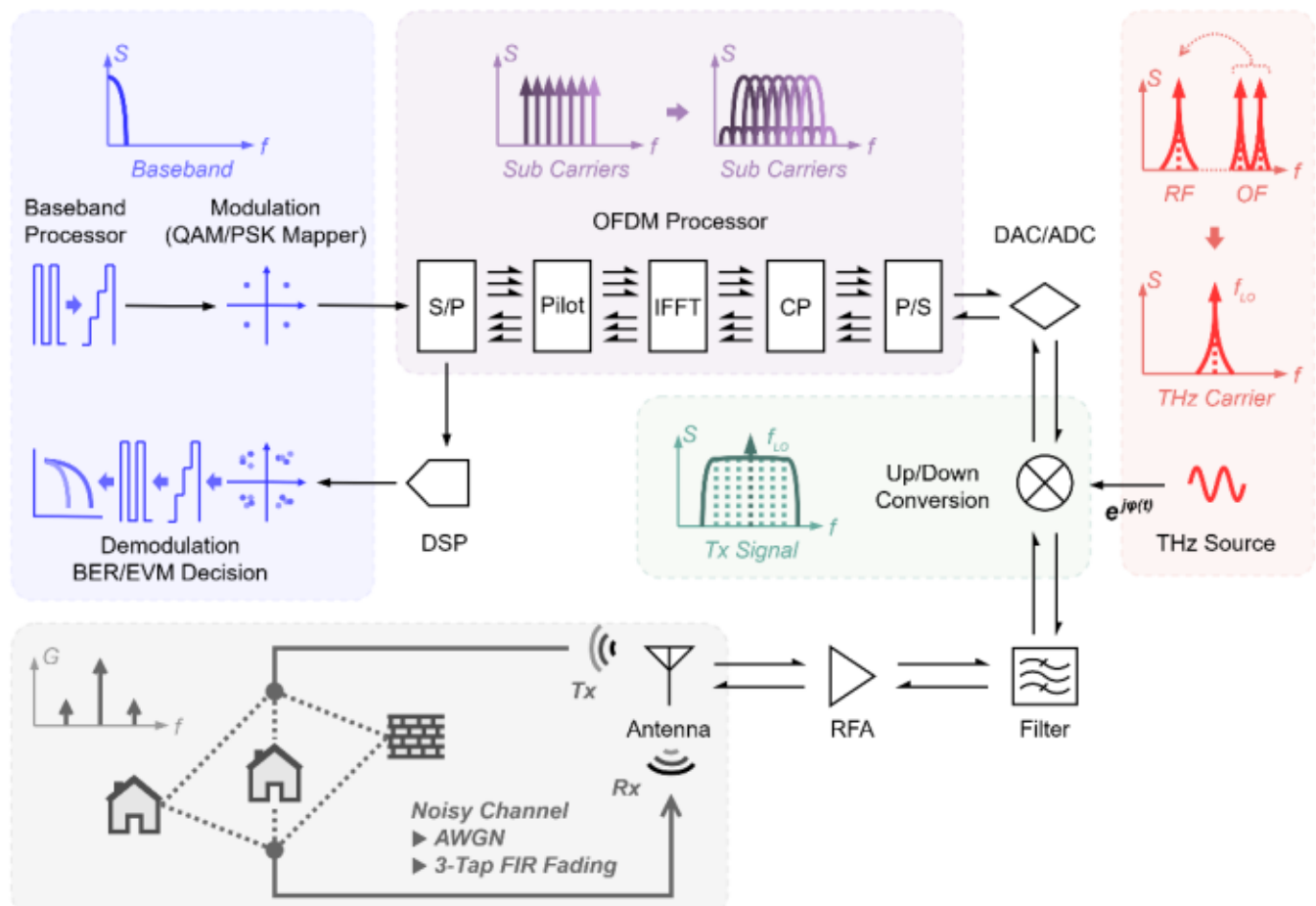

Fig. 2. Processing chain of the noisy THz-link simulation model.

### B. Phase Noise Modeling and Signal Processing

The phase noise modeling and data processing workflow is illustrated in Fig. 3. To ensure practical relevance, phase noise is first extracted from experimentally measured PSD masks. The PSD data is interpolated to match the simulation grid and converted to a linear frequency scale $L(f)$. To emulate the inherent randomness of phase noise, uniformly distributed random phases $\theta(f)$ within the range of $[-\pi, +\pi]$ are added [22].

$$O(f) = \begin{cases} \sqrt{\dfrac{L(f)}{2}}\, e^{j\theta(f)} \quad , & f > 0 \\ \sqrt{\dfrac{L(-f)}{2}}\, e^{-j\theta(f)}, & f < 0 \\ 1 \quad , & f = 0 \end{cases} \tag{2}$$

IFFT is then applied to obtain the time-domain phase jitter $\varphi(t)$, which is subsequently imposed on the baseband symbols $u(n)$ during the up-conversion process, resampled time sequence $t$ to symbol index $n$ according to the system sampling rate.

$$\varphi(t) = Re\{\mathcal{F}^{-1}[O(f)]\} \tag{3}$$

$$s(t) = u(n) \cdot e^{j\varphi(n)}, \qquad t \equiv n \tag{4}$$

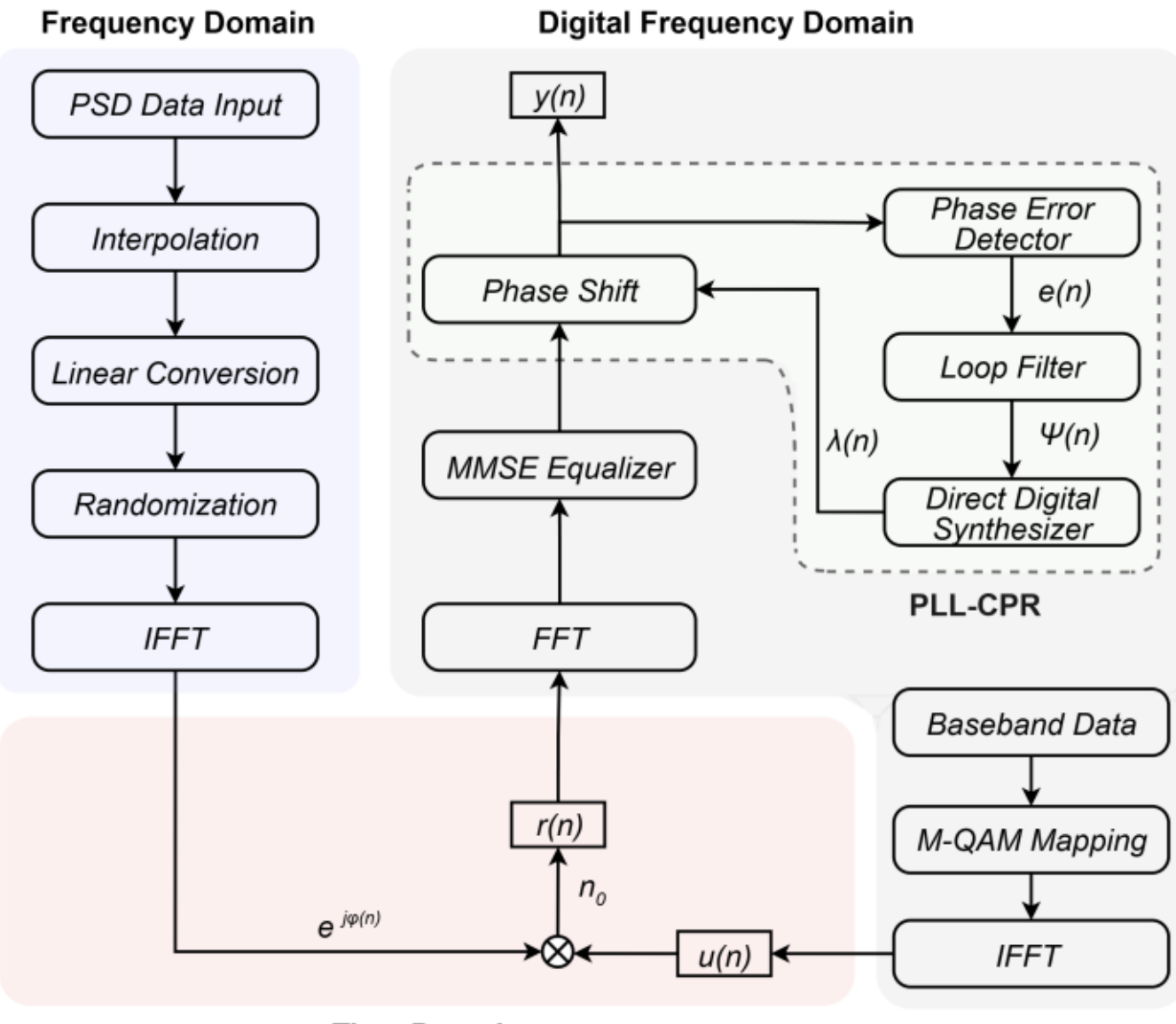

Fig. 3. Processing chain of phase noise modeling and error estimation.

Table 1 summarizes the representative low-phase-noise local oscillators (LOs). These include a standard CMOS oscillator defined by the 3GPP 5G NR protocols [6], a low-noise high-frequency 28-nm CMOS oscillator [23], an ultra-robust analog signal generator (ASG, Keysight E8257D) [11], external cavity laser (ECL) diodes [8], and a silicon nitride ($Si_3N_4$) MRR [24]. The standard deviations (STD, σ) of their phase noise contributions are statistically evaluated within a 100 Hz ~ 100 MHz offset bandwidth, scaled to a 300-GHz carrier frequency. The phase noise characteristics of these LOs are reconstructed and illustrated in Fig. 4. High-frequency instantaneous noise $\varphi_{inst}$ is extracted by a smooth filter and highlighted. As shown in Fig. 4 (a), different LOs exhibit distinct phase noise levels. The standard CMOS oscillator suffers from significant low- and mid-frequency phase noise, while the 28-nm CMOS oscillator features a notably elevated high-frequency noise floor. In contrast, the $Si_3N_4$ MRR exhibits the lowest overall phase noise, with particularly suppressed instantaneous noise components. These observations are consistent with the noise behaviors defined by the PSD spectra depicted in Fig. 4 (b).

TABLE I. PHASE NOISE OF LOW-NOISE OSCILLATORS

| *Osc.* | $f_{LO}$ (GHz) | BW: $10^2$ ~ $10^8$ Hz | | *Ref.* |
|---|---|---|---|---|
| | | *Full σ* | *Inst. σ* | |
| Standard CMOS | 29.55 | 0.8683 | 0.2602 | [6] |
| 55-nm BiCMOS | 9.96 | 13.8762 | 0.0310 | [25] |
| 28-nm CMOS | 300 | 0.1540 | 0.1001 | [23] |
| 14-nm FinFET | 6.45 | 0.0764 | 0.0564 | [26] |
| ASG E8257D | 50 | 0.0148 | 0.0115 | [11] |
| SIL $Si_3N_4$ | 50 | 0.8976 | 0.0120 | [11] |
| ECL | 350 | 0.0252 | 0.0212 | [8] |
| SBS Fiber Laser | 137.5 | 0.4295 | 0.0108 | [27] |
| 2P-OFD $Si_3N_4$ | 20 | 0.0071 | 0.0058 | [24] |

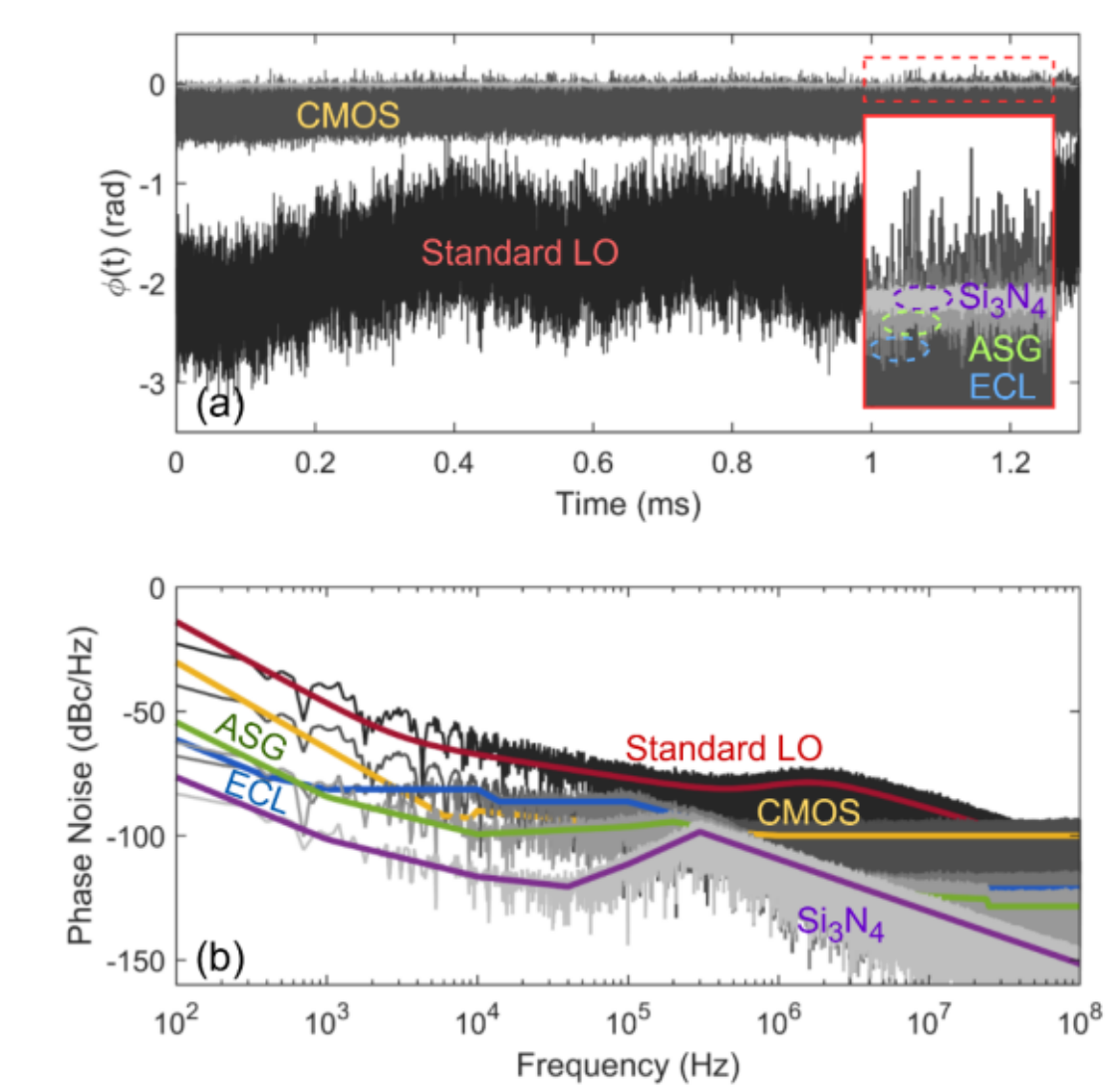

Fig. 4. Phase noise modeled from experimental data: (a) time-domain phase noise; (b) PSD in frequency domain.

Following phase noise introduction, the transmit signal $s(n)$ propagates through a standard AWGN channel introducing additive noise $n_0$, with simplified frequency-selective fading. These channel impairments frequency-dependent delays, and amplitude distortion to the received signal $r(n)$. To focus on impact of phase noise, compensation for less related channel effects is simplified. At the receiver, $r(n)$ undergoes DSP include timing synchronization, equalization and carrier phase recovery (CPR). A pilot-assisted minimum mean square error (MMSE) equalizer estimates the channel response to mitigate distortions from fading and additive noise.

$$W_{MMSE} = \frac{H^H}{H^H H + N_0 I} = \frac{h^*}{|h|^2 + 1/SNR} \tag{5}$$

Subsequent PPL-based CPR module compensates for residual phase errors, where the processing chain of the loop depends on a phase shift, a phase error detector $e(n)$, a loop filter $\psi(n)$ and a direct digital synthesizer $\lambda(n)$.

$$e(n) = sgn(Re\{r(n)\} \times Im\{r(n)\} - sgn(Im\{r(n)\}) \times Re\{r(n)\}) \tag{6}$$

$$\psi(n) = \psi(n-1) + e(n) \cdot \frac{2\theta^2}{1 + 2\zeta\theta + \theta^2}, \theta = \frac{BT}{\zeta + \frac{1}{4\zeta}} \tag{7}$$

$$\lambda(n) = \lambda(n-1) + \psi(n-1) + \frac{e(n-1) \cdot 2\zeta\theta}{1 + 2\zeta\theta + \theta^2} \tag{8}$$

$$y(n) = r(n) \cdot e^{j\lambda(n)} \tag{9}$$

$B$ is the normalized loop bandwidth; $T$ denotes phase lock delay; and $\zeta$ gives the damping factor. The corrected symbol $y(n)$ is then used for symbol-level error evaluation.

### C. Pilot and Channel Fading Configuration

The symbol pattern is illustrated in Fig. 5. Pilot tones are inserted in a comb-type arrangement, within every 16 sub-carriers, resulting in 128 pilot number and an overhead of 6.25%. The ratio is adjusted to explore the trade-off between pilot overhead and system robustness. The comb-pattern is particularly for mitigating the distortion introduced by PM-FM-AM converting noise, which manifests as ICI in OFDM transmission. A CP of 16 sub-carriers is added to each symbol to combat ISI. This length is sufficient for a small number of delayed propagation paths. To maintain simulation tractability while preserving physical relevance, the channel fading is modeled using a 3-tap FIR filter. This model captures the effects of three distinct propagation paths with varying delays and attenuations (or gains), which is representative of typical indoor or short-range line-of-sight (LoS) wireless environments, such as chip-to-chip, room-scale, or device-to-device (D2D) THz communication scenarios [28, 29]. This setup enables clearer isolation and analysis of phase noise induced degradation via minimizing variable channel conditions.

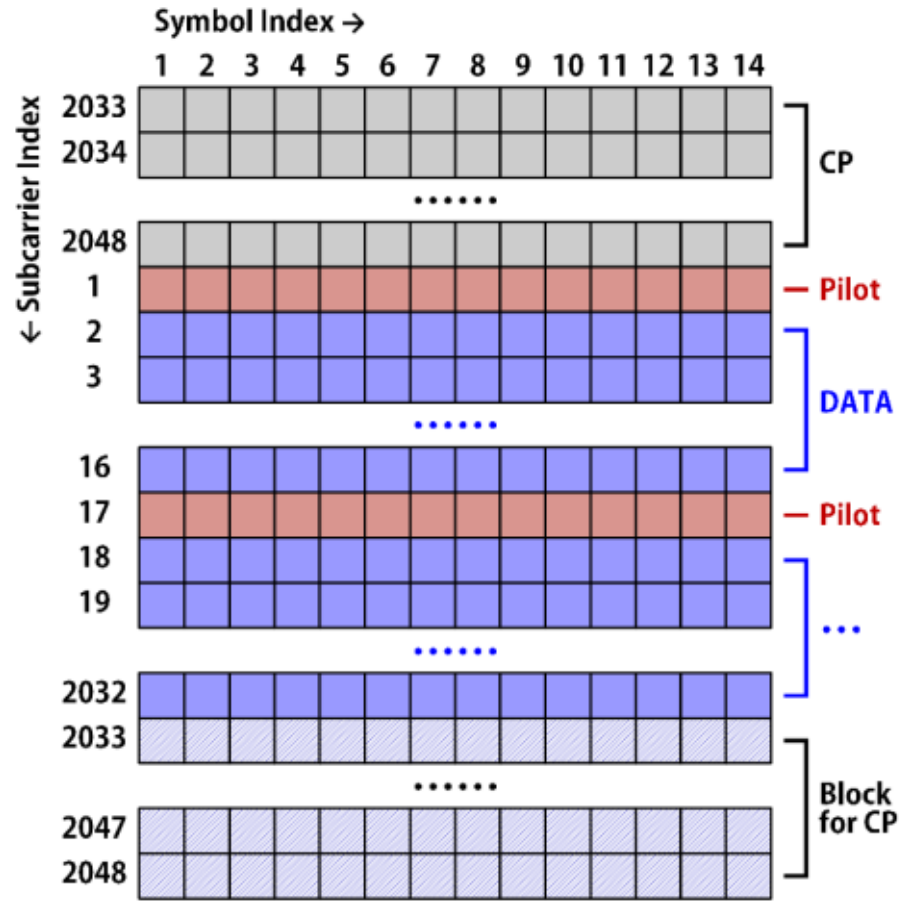

Fig. 5. Symbol pattern after channel encoding with pilot and CP signals.

## III. Results and Discussion

### A. 64-QAM 2048-OFDM Terahertz Transmission Driven by Different Local Oscillators

Figure 6 presents the simulated constellation diagrams driven by proposed THz models: a standard CMOS oscillator [6], photonic mixing external cavity laser (ECL) diodes [8], an ultra-low-noise ASG [11], and a stabilized $Si_3N_4$ MRR µ-comb [24]. Three stages of received symbols are illustrated at a SNR of 40 dB: (1) without DSP, (2) after equalization, and (3) after CPR. As the STD of phase noise decreases, the symbol clouds become more compact, indicating improved signal recovery. In this dense multi-carrier configuration with 64-QAM modulation, CMOS oscillator that is commonly adopted in 5G NR, fails to support the transmission. Severely smeared constellations with an RMS EVM of 91.8%. ECL also suffers from considerable degradation with 35.7% EVM. In contrast, both ASG and MRR schemes show minimal global rotation. This is attributed to their lower low-frequency phase noise, which suppresses common phase error (CPE). Particularly for $Si_3N_4$ MRR, benefits from its exceptionally low noise floor ($\sigma_{inst}$ = 0.0058), the lowest EVM of 4.7% is realized, outperforming 5.9% of ASG. According to our prior results in single-carrier 64-QAM links, an EVM threshold of 5.12% corresponds to a 3σ decision criterion for error-free transmission with no FEC [30]. Therefore, under the same noisy level, $Si_3N_4$ MRR satisfies 3σ criterion while ASG cannot. This highlights the critical role of instantaneous component $\varphi_{inst}$ in determining the performance ceiling of dense OFDM systems.

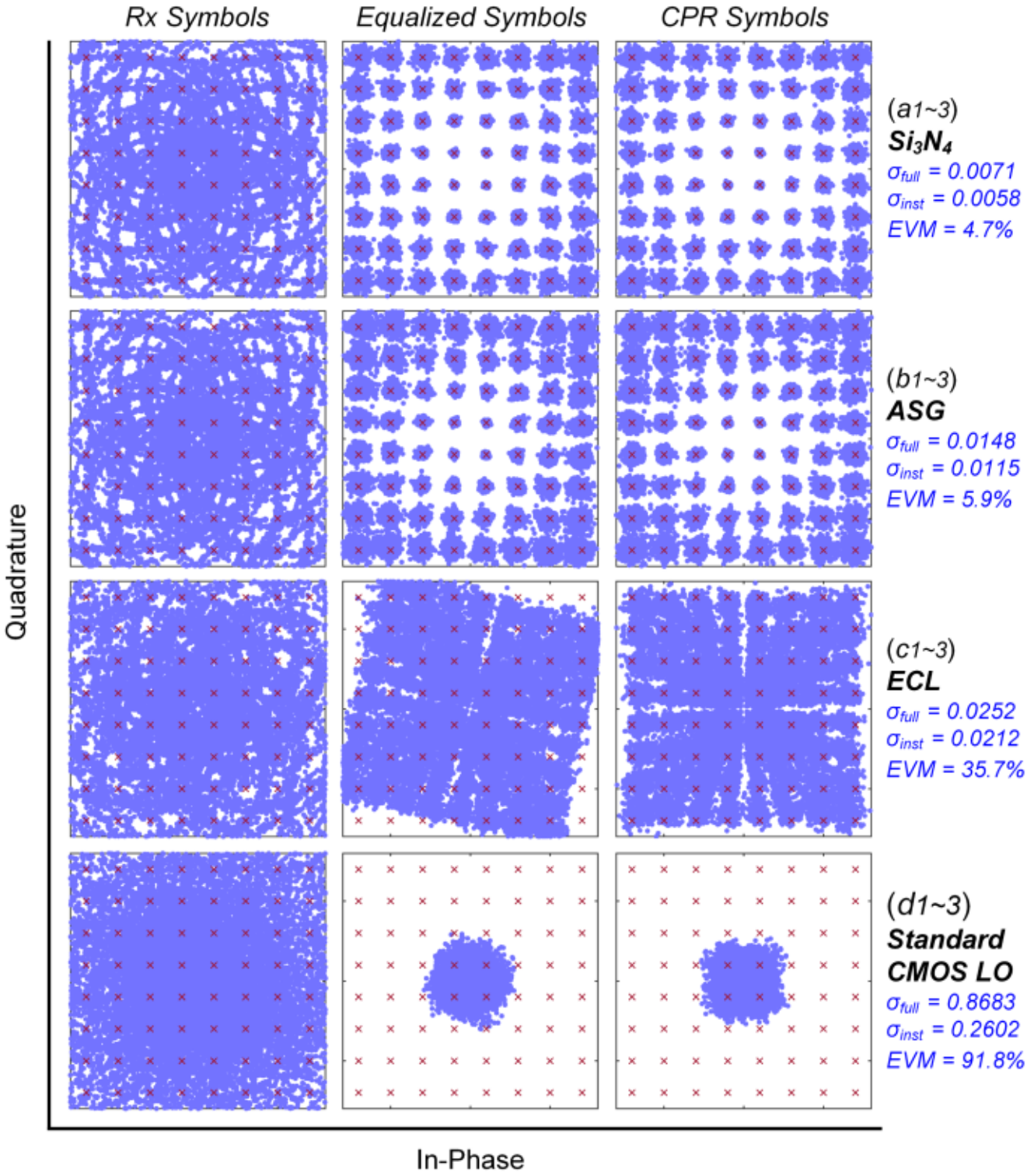

Fig. 6. 64-QAM constellation diagrams driven by different local oscillators before and after equalization and CPR processing.

### B. Instantaneous Phase Noise Constrains Error Ceiling

To further investigate the performance ceiling imposed by instantaneous phase noise, we evaluate the system behavior using the $Si_3N_4$ MRR [24] as the THz source under varying optical SNR. As shown in Fig. 7(a), constellation diagrams are plotted at SNR levels of 30 dB, 40 dB, and 50 dB. Increasing SNR leads to a rapid improvement in constellation clarity and a corresponding fall in EVM as expected. However, beyond an SNR of approximately 50 dB, the rate of EVM reduction diminishes significantly, eventually converging to a floor value of 2.71% as depicted in Fig. 7(b). This saturation behavior indicates that the residual distortion from instantaneous phase noise cannot be further mitigated by increasing SNR alone. It causes irreducible ICI, particularly affecting outer constellation points, which suffer more pronounced scattering. Under such white phase noise, boosting SNR indefinitely yields no meaningful improvement. According to 3σ decision criterion [30], an SNR of at least 38 dB is required to ensure that 99.7% of received symbols fall within the correct decision boundary. This 3σ radius is computed based on the RMS amplitude of the M-ary QAM constellation, and defines the region where symbol detection can be considered statistically reliable without FEC.

$$r_{3\sigma} = 3 \times EVM_{RMS}\sqrt{P_M} = EVM_{RMS}\sqrt{6(M-1)} \tag{10}$$

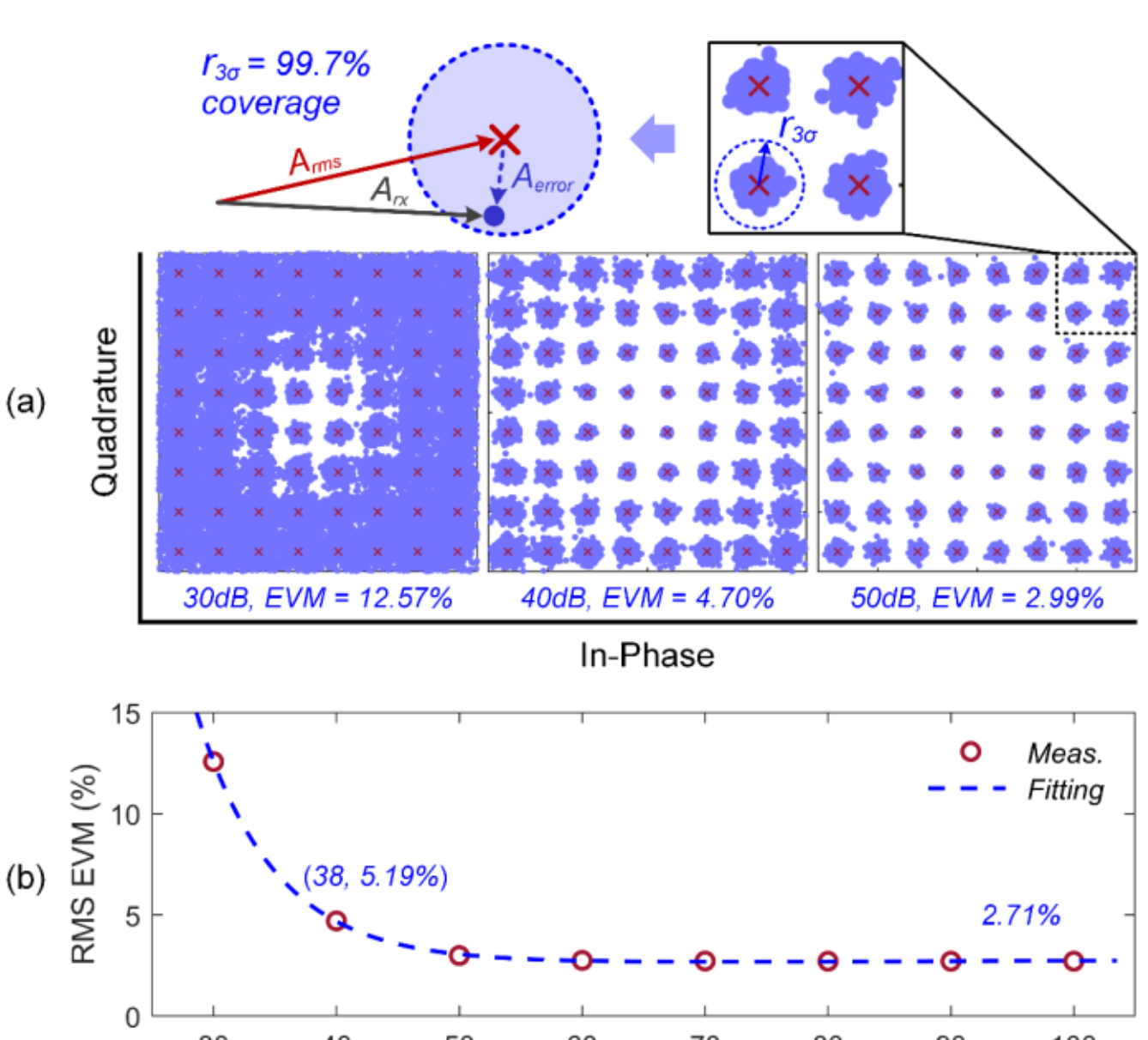

Fig. 7. (a) Phase error recovered constellations with 30 dB, 40 dB and 50 dB SNR using $Si_3N_4$ MRR (insert: 3σ-error estimation); (b) EVM versus SNR.

### C. Low Instantaneous Phase Noise Reduces Pilot Overhead

The impact of pilot overhead on system performance is further evaluated using the same $Si_3N_4$ MRR [24] as the LO. Figure 8 (a) shows the evolution of constellation diagrams as the number of pilot sub-carriers increases. With only 8 pilots, severe distortion is observed, resulting in an EVM of 15.64%. As the pilot count increases to 64 and 512, the constellations become significantly more compact, and the EVM improves to 2.88% and 2.23%, respectively. Figure 8(b) summarizes the EVM performance as a function of pilot density, comparing the MRR [24] with the ASG [11]. Both systems exhibit a similar asymptotic decay in EVM as the number of pilots increases. For the MRR-driven system, the EVM converges to a floor of 1.75% when using 64 pilots (0.78% overhead). In contrast, the ASG system requires at least 128 pilots (1.56% overhead) to reach a higher EVM floor of 2.98%. This comparison reveals that even a relatively small difference in instantaneous phase noise STD (Δσ ≈ 0.006) leads to a double increase in pilot overhead requirement. The MRR μ-comb lower noise floor allows for significantly reduced pilot density without sacrificing signal fidelity, offering clear advantages in spectral efficiency. These results highlight that for high-order OFDM systems, minimizing instantaneous phase noise not only improves signal recovery quality but also reduces protocol complexity by greatly lowering the required pilot burden. More importantly, this reinforces the advantages of MRR μ-comb for ultra-low-pilot-overhead OFDM links beyond 64-QAM, which directly translates into higher spectral efficiency, bit rates and energy utilization.

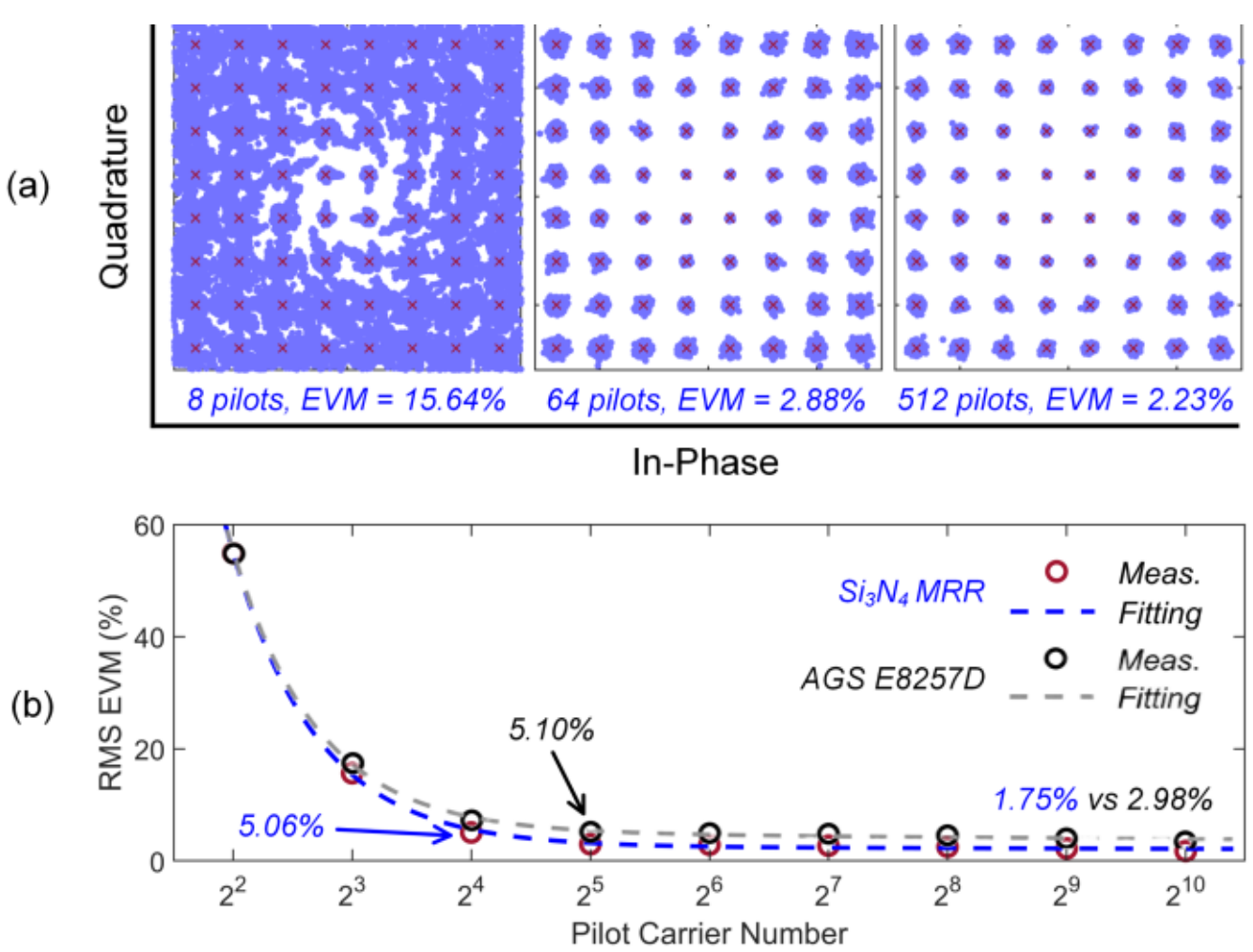

Fig. 8. (a) Phase error recovered constellations respectively encoded with 8, 64 and 512 pilots using $Si_3N_4$ MRR; (b) EVM versus pilot overhead.

## IV. Conclusion

In this work, we quantitatively investigated the impact of phase noise, particularly of instantaneous component, on the performance of high-order OFDM THz wireless systems. By modeling realistic phase noise spectra extracted from representative THz sources, we demonstrated that even under high SNR conditions, irrecoverable white phase noise induces impairments persist and sets a fundamental limit on achievable ceiling. Through constellation analysis and 3σ error probability estimation, we established that MRR optical frequency combs provide superior phase noise tolerance, enabling error-free 64-QAM 2048-OFDM transmission with significantly lower pilot overhead. Compared with low-noise analog and electronic LOs, MRR μ-comb achieved a lower EVM floor (1.75%) with only 0.78% pilot overhead, due to its ultralow noise floor. The results reveal that minimizing instantaneous phase noise not only improves signal fidelity but also allows for substantial reduction in DSP complexity and protocol overhead. We suggest that the adoption of photonic MRR-based LOs is crucial to unlocking practical, high-capacity, low-pilot-overhead OFDM THz links beyond 64-QAM.

## Acknowledgment

This work was supported by JST, CRONOS, Japan Grant Number JPMJCS24N7.